# Quantum entanglement dynamics of the three-qubit $W_\zeta$ quantum state coupled to spin chain with ternary interaction

S. M. Moosavi Khansari[1]
*Department of Physics, Faculty of Basic Sciences, Ayatollah Boroujerdi University, Boroujerd, IRAN*
F. Kazemi Hasanvand[2]
*Department of Physics, Faculty of Basic Sciences, Ayatollah Boroujerdi University, Boroujerd, IRAN*

In this study, we explore the dynamics of quantum entanglement using the negativity criterion for the $W_\zeta$ quantum state. We investigate changes in negativity in terms of anisotropy parameters, $\gamma$, the strength of the external magnetic field applied to the spin chain, $\eta$, the triple interaction strength, $\alpha$. We examine how these parameters affect the entanglement properties of the system and discuss the implications for quantum information processing and quantum communication protocols. By analyzing the negativity of the $W_\zeta$ state under different conditions, we gain insights into the behavior of entanglement in complex quantum systems. Our results shed light on the intricate interplay between various factors that influence quantum entanglement and provide a foundation for further investigations in this field of research.

## I Introduction

Quantum entanglement, as a key aspect of quantum theory, has garnered significant interest in recent years. Its nonlocal correlations between quantum systems are crucial in quantum information processing. Quantum entanglement is the bases of several tasks in quantum information and computation [1-8]. However, real-world non-local entanglement is often sensitive to decoherence, which refers to the loss of quantum coherence as a system interacts with its environment. This intricate quantum phenomenon poses challenges, leading to diverse studies using various models [9-16]. Presently, there is a growing focus on investigating decoherence within systems featuring a spin environment exhibiting a quantum phase transition at zero temperature induced by pure quantum fluctuations [17-24]. In this context, understanding entanglement as a geometric phase [25] and analyzing fidelity [26] are fundamental for studying quantum phase transitions. In prior research, attention was primarily given to the nearest spin-spin interactions in spin chains. However, in practical applications, it becomes essential to delve into quantum spin models that accurately account for not only nearest neighbour interactions but also consider nearest neighbour spin exchange models or multiple spin exchange models [27, 28, 29, 30, 31]. Research in this area has thus expanded to encompass more sophisticated quantum spin models that can better capture the intricacies of real-world systems. By considering not only nearest neighbor interactions but also incorporating nearest neighbor spin exchange models or even multiple spin exchange models, researchers hope to gain deeper insights into the behavior of these quantum systems. Studies involving these more complex models, such as those proposed by Tsvelik et al. [23], Frahm et al. [24], Zvyagin [25], and others, are shedding light on the effects of various interactions on quantum entanglement and decoherence in spin environments. By exploring these richer models, scientists aim to improve our understanding of quantum phase transitions and pave the way for more advanced applications of quantum information processing technologies [32-36].

---
[1] Email of the corresponding author: m.moosavikhansari@abru.ac.ir
[2] E-mail: fa_kazemi270@yahoo.com

This study investigates entanglement dynamics in a qubit system embedded in a spin chain environment. The organization of the rest of this paper is as follows: In Sec. II, we introduces the model and derives the time-dependent density operator. In Section III, the $W_\xi$ state is introduced, followed by the derivation of its time-dependent density matrix and the calculation of measures from Section II. Section IV is devoted to results and discussion. Finally, section V is devoted to conclusions.

Recent experiments have convincingly demonstrated that entanglement is not limited to micro-scale systems, but can indeed manifest in a variety of macroscopic and strongly correlated solids. Within this interesting category, certain magnetic compounds, particularly $Na_2Cu_5Si_4O_{14}$ and $MgMnB_2O_5$, have shown to function effectively as reservoirs of entanglement. These specific magnetic materials lend themselves to extensive analysis through the application of variational spin-chain models, which provide a structured approach to examining and understanding the intricacies of entanglement in these complex systems. Theoretical investigations within this field have placed significant emphasis not only on traditional nearest-neighbor spin-spin interactions, which have been the focus of much previous work, but also on the next-nearest-neighbor interactions and the dynamics involved in multiple spin-exchange models. By incorporating these additional interactions, the resulting frameworks offer a more accurate and comprehensive representation of real-world scenarios compared to simpler models that only account for nearest-neighbor interactions. For example, the inclusion of additional terms that involve multisite interaction operators has proven to be critically important for adequately describing a wide range of physical systems, including but not limited to solid $^3$He. This particular case illustrates the necessity of considering these more complex interactions, which have largely been neglected in past research. Indeed, previous studies concerning decoherence driven by the spin environment have often overlooked these crucial interactions, despite the fact that they frequently occur in actual spin systems. Thus, this work aims to deeply investigate how multisite interactions exert their influence on the disentanglement process, contributing valuable insights into the dynamics of entanglement in complex magnetic materials [37-42].

## II  Hamiltonian and the time dependent density matrix

In this article, we express the Hamiltonian of the entire composite system with the following relation

$$H = H_0 + H_I \qquad (1)$$

where

$$H_0 = -\sum_{k=-M}^{M} \alpha\left(\sigma_{k-1}^x \sigma_k^z \sigma_{k+1}^y - \sigma_{k-1}^y \sigma_k^z \sigma_{k+1}^x\right)$$
$$-\sum_{k=-M}^{M} \left(\frac{1}{2}(\gamma+1)\sigma_k^x \sigma_{k+1}^x + \frac{1}{2}(1-\gamma)\sigma_k^y \sigma_{k+1}^y + \eta \sigma_k^z\right) \qquad (2)$$

expresses the spin chain Hamiltonian of Inthe environment with the $XZY - YZX$ triplet interaction and

$$H_I = \sum_{k=-M}^{M} (-g_A \sigma_A^z - g_B \sigma_B^z - g_C \sigma_C^z) \sigma_k^z \tag{3}$$

is the Hamiltonian of the interaction between the three-qubit system and the spin chain of the environment. In these relations, $\sigma_A^z$, $\sigma_B^z$, $\sigma_C^z$ and $\sigma_k^x$, $\sigma_k^y$ and $\sigma_k^z$ are Pauli operators that describe the three qubits and the spin chain of the environment. The parameter $M = (N - 1)/2$ is defined for odd $N$, where $N$ is the number of particles in the spin chain. The parameter $\gamma$ determines the anisotropy in the in-plane interaction, while $\eta$ expresses the strength of the externally applied transverse field to the environment. The parameters $g_A$, $g_B$, and $g_C$ represent the coupling strength between the three-qubit system and the spin chain of the environment. The parameter $\alpha$ describes the strength of the $XZY - YZX$ triplet interaction. This establishes the following quantum commutator

$$[g_A \sigma_A^z + g_B \sigma_B^z + g_C \sigma_C^z, \sigma_k^{x,y,z}] = 0 \tag{4}$$

Therefore, the total Hamiltonian can be expressed as follows

$$H = \sum_{\mu=1}^{8} |\phi_\mu\rangle\langle\phi_\mu| \otimes H_E^{\lambda_\mu} \tag{5}$$

where $|\phi_\mu\rangle$ represents the $\mu$ th eigenstate of the operator $(g_A \sigma_A^z + g_B \sigma_B^z + g_C \sigma_C^z)$ and $|000\rangle,\ldots,|111\rangle$ corresponding to the $\mu$th eigenvalue of $g_\mu$ are taken into account. Additionally, $\lambda_\mu = \eta + g_\mu$, and these values are defined by the following expressions

$$\lambda_1 = g_A + g_B + g_C + \eta, \quad \lambda_2 = g_A + g_B - g_C + \eta \tag{6}$$

$$\lambda_3 = g_A - g_B + g_C + \eta, \quad \lambda_4 = g_A - g_B - g_C + \eta \tag{7}$$

$$\lambda_5 = -g_A + g_B + g_C + \eta, \quad \lambda_6 = -g_A + g_B - g_C + \eta \tag{8}$$

$$\lambda_7 = -g_A - g_B + g_C + \eta, \quad \lambda_8 = -g_A - g_B - g_C + \eta \tag{9}$$

To determine the entanglement dynamics of the target three-qubit system, acquiring the time evolution operator $U(t) = exp(-iHt)$ is essential. Given the initial state of the system, $\rho(0)$, we can derive the quantum state's time evolution using $\rho(t) = U(t)\rho(0)U(t)^\dagger$. Utilizing the Jordan-Wigner transformation, which converts the spin system into a quasi-Fermi system

$$\sigma_k^z = 1 - 2c_k^+ c_k, \quad \sigma_k^+ = [\prod_{m<k} (1 - 2c_m^+ c_m)]c_k, \quad \sigma_k^- = [\prod_{m<k} (1 - 2c_m^+ c_m)]c_k^\dagger \tag{10}$$

and the Fourier transform

$$d_\ell = \frac{1}{\sqrt{N}} \sum_k c_k exp(-2\pi ik\ell/N) \tag{11}$$

and Bogolyubov transformation

$$b_{\ell,\lambda_\mu} = d_\ell cos\frac{\theta_\ell^{\lambda_\mu}}{2} - id_{-\ell}^\dagger sin\frac{\theta_\ell^{\lambda_\mu}}{2} \tag{12}$$

where

$$tan\left(\theta_\ell^{\lambda_\mu}\right) = \frac{\gamma sin\left(\frac{2\pi\ell}{N}\right)}{\left(\lambda_\mu - cos\left(\frac{2\pi\ell}{N}\right)\right)} \tag{13}$$

the $H_E^{\lambda_\mu}$ Hamiltonian can be accurately diagonalized as [26]
$$H_E^{\lambda_\mu} = \sum_{k=-M}^{M} \xi_k^{\lambda_\mu} \left(b_{k,\lambda_\mu}^\dagger b_{k,\lambda_\mu} - \frac{1}{2}\right) \tag{14}$$

the energy spectrum can be expressed as
$$\xi_k^{\lambda_\mu} = 2\alpha\sin\left(\frac{4\pi k}{N}\right) + 2\sqrt{\gamma^2\sin^2\left(\frac{2\pi k}{N}\right) + \left(\lambda_\mu - \cos\left(\frac{2\pi k}{N}\right)\right)^2} \tag{15}$$

With these interpretations, we can explore the dynamic evolution of the three-qubit system. We assume that the three-qubit system and the spin chain of the environment initially have a density matrix in the form of the following tensor product
$$\rho(0) = \rho_{ABC}(0) \otimes \rho_E(0) \tag{16}$$

where $\rho_{ABC}(0)$ and $\rho_E(0)$ represent the initial density matrix of the three-qubit system and the density matrix of the spin chain of the environment, respectively. Now, we assume that the three-qubit system is initially in the state $\rho_{ABC}(0) = \sum_{\mu,\nu} |\phi_\mu\rangle\langle\phi_\nu|$, while the spin chain of the environment is initially in thermal equilibrium state as follows
$$\rho_E(0) = \frac{1}{Z}\exp\left(-\beta \sum_k \xi_k b_k^\dagger b_k\right) \tag{17}$$

where $\beta = \frac{1}{kT}$ and we can express
$$\xi_k = 2\alpha\sin\left(\frac{4\pi k}{N}\right) + 2\sqrt{\gamma^2\sin^2\left(\frac{2\pi k}{N}\right) + \left(\eta - \cos\left(\frac{2\pi k}{N}\right)\right)^2} \tag{18}$$

and
$$Z = Tr\left[\exp\left(-\beta \sum_k \xi_k b_k^\dagger b_k\right)\right] \tag{19}$$

where Z represents the partition function of the spin chain environment. Upon formulating the time evolution of the entire system and tracing the environment spin chain variable, the reduced density matrix $\rho_{ABC}(t) = Tr_E[U(t)\rho(0)U(t)^\dagger]$ for the three-qubit system can be derived using the subsequent equation
$$\rho_{ABC}(t) = \sum_{\mu,\nu} \rho_{\mu,\nu}(t)|\phi_\mu\rangle\langle\phi_\nu| \tag{20}$$

where $\rho_{\mu,\nu}(t)$ is written as follows
$$\rho_{\mu,\nu}(t) = \rho_{\mu,\nu}(0)F_{\mu,\nu}(t) \tag{21}$$

in this mathematical relation, $F_{\mu,\nu}(t)$ is defined as follows
$$F_{\mu,\nu}(t) = \prod_{k>0} \frac{e^{it(\xi_{k,\lambda_\mu} - \xi_{k,\lambda_\nu})}}{Z_k}\left(\mathcal{A}_k^{\lambda_\mu,\lambda_\nu} + 2e^{-\beta\xi_{k,\eta} + it(\xi_{k,\lambda_\nu} - \xi_{k,\lambda_\mu})} + e^{-2\beta\xi_{k,\eta}}\mathcal{B}_k^{\lambda_\mu,\lambda_\nu}\right) \tag{22}$$

where $\mathcal{A}_k^{\lambda_\mu,\lambda_\nu}$ and $\mathcal{B}_k^{\lambda_\mu,\lambda_\nu}$ are defined as follows
$$\mathcal{A}_k^{\lambda_\mu,\lambda_\nu} = \left(1 - e^{2it\xi_{k,\lambda_\mu}}\right)\left(1 - e^{-2it\xi_{k,\lambda_\nu}}\right)\sin\left(\frac{1}{2}(\theta_{k,\lambda_\mu} - \theta_{k,\eta})\right)\sin\left(\frac{1}{2}(\theta_{k,\lambda_\nu} - \theta_{k,\eta})\right)$$

$$\cos\left(\tfrac{1}{2}\left(\theta_{k,\lambda_\mu}-\theta_{k,\lambda_\nu}\right)\right)-\left(\left(1-e^{2it\xi_{k,\lambda_\mu}}\right)\sin^2\left(\tfrac{1}{2}\left(\theta_{k,\lambda_\mu}-\theta_{k,\eta}\right)\right)\right)$$
$$-\left(1-e^{-2it\xi_{k,\lambda_\nu}}\right)\sin^2\left(\tfrac{1}{2}\left(\theta_{k,\lambda_\nu}-\theta_{k,\eta}\right)\right)+1 \tag{23}$$

$$\mathcal{B}_k^{\lambda_\mu,\lambda_\nu} = \left(1-e^{2it\xi_{k,\lambda_\mu}}\right)\left(1-e^{-2it\xi_{k,\lambda_\nu}}\right)\cos\left(\tfrac{1}{2}\left(\theta_{k,\lambda_\mu}-\theta_{k,\eta}\right)\right)\cos\left(\tfrac{1}{2}\left(\theta_{k,\lambda_\nu}-\theta_{k,\eta}\right)\right)$$
$$\cos\left(\tfrac{1}{2}\left(\theta_{k,\lambda_\mu}-\theta_{k,\lambda_\nu}\right)\right)-\left(\left(1-e^{2it\xi_{k,\lambda_\mu}}\right)\cos^2\left(\tfrac{1}{2}\left(\theta_{k,\lambda_\mu}-\theta_{k,\eta}\right)\right)\right)$$
$$-\left(1-e^{-2it\xi_{k,\lambda_\nu}}\right)\cos^2\left(\tfrac{1}{2}\left(\theta_{k,\lambda_\nu}-\theta_{k,\eta}\right)\right)+1 \tag{24}$$

With these interpretations, we have acquired the reduced density matrix $\rho_{ABC}(t)$, that can be utilized to examine the entanglement dynamics of the target quantum state for the three-qubit system.

We employ the negativity measure to compute entanglement. For a quantum state with density matrix $\rho$, negativity is defined as [27]

$$N(\rho) = \frac{||\rho^{T_j}||-1}{2} \tag{25}$$

In this context, $\rho^{T_j}$ represents the partial transpose of the density matrix $\rho$ with respect to the $j$ component. If $N > 0$, this indicates that the quantum state possesses entanglement; conversely, if $N = 0$, the state is classified as separable, which means it is not entangled in any way. The degree of negativity is crucial as a higher negativity value signifies a stronger level of entanglement present in the system. When it comes to analyzing complex quantum systems, researchers often employ numerical methods such as Monte Carlo techniques or various optimization algorithms. These methods are particularly useful for estimating negativity, especially when working with large and intricate density matrices that can be challenging to handle. By using these advanced techniques, we can systematically quantify the phenomenon of quantum entanglement. Furthermore, they assist in deepening our understanding of the intricate quantum correlations that exist within these systems. In our research, we specifically utilize the negativity measure as a primary tool to compute and analyze entanglement effectively.

In this study, the density matrix is denoted as $\rho_{ABC}(t)$ and $j$ can refer to the $A$, $B$, or $C$ subsystems. Hence, the negativity of the state mentioned is defined as

$$N_{A-BC}(\rho(t)) = \frac{||\rho(t)^{T_A}||-1}{2} \tag{26}$$

or

$$N_{B-CA}(\rho(t)) = \frac{||\rho(t)^{T_B}||-1}{2} \tag{27}$$

or

$$N_{C-AB}(\rho(t)) = \frac{||\rho(t)^{T_C}||-1}{2} \tag{28}$$

### III  The $W_\zeta$ state serves as the initial state for the three-qubit system

In this section, we define the state $W_\zeta$ as the initial state of the system as follows:
$$|W_\zeta\rangle = \frac{1}{\sqrt{2\zeta+2}}(|100\rangle + e^{i\phi}\sqrt{\zeta}|010\rangle + e^{i\delta}\sqrt{\zeta+1}|001\rangle) \qquad (29)$$

where $\zeta$ is a natural number and $\delta$ and $\phi$ are real numbers.
The initial density matrix of the system corresponding to this state is defined as follows:

$$[\rho_s(0)]_{W_\zeta} = \begin{pmatrix} 0 & 0 & 0 & 0 & 0 & 0 & 0 & 0 \\ 0 & \frac{1}{2} & \frac{\zeta e^{i(\delta-\phi)}}{2\sqrt{\zeta(\zeta+1)}} & 0 & \frac{e^{i\delta}}{2\sqrt{\zeta+1}} & 0 & 0 & 0 \\ 0 & \frac{\zeta e^{-i(\delta-\phi)}}{2\sqrt{\zeta(\zeta+1)}} & \frac{\zeta}{2\zeta+2} & 0 & \frac{\sqrt{\zeta}e^{i\phi}}{2\zeta+2} & 0 & 0 & 0 \\ 0 & 0 & 0 & 0 & 0 & 0 & 0 & 0 \\ 0 & \frac{e^{-i\delta}}{2\sqrt{\zeta+1}} & \frac{\sqrt{\zeta}e^{-i\phi}}{2\zeta+2} & 0 & \frac{1}{2\zeta+2} & 0 & 0 & 0 \\ 0 & 0 & 0 & 0 & 0 & 0 & 0 & 0 \\ 0 & 0 & 0 & 0 & 0 & 0 & 0 & 0 \\ 0 & 0 & 0 & 0 & 0 & 0 & 0 & 0 \end{pmatrix} \qquad (30)$$

After some algebraic calculations, the density matrix of the system following interaction with the environment for this state can be expressed as a matrix and in relation to eight bases of the corresponding three qubits as shown:

$$[\rho_s(t)]_{W_\zeta} =$$
$$\begin{pmatrix} 0 & 0 & 0 & 0 & 0 & 0 & 0 & 0 \\ 0 & \frac{1}{2} & \frac{\zeta F_{2,3} e^{i(\delta-\phi)}}{2\sqrt{\zeta(\zeta+1)}} & 0 & \frac{e^{i\delta} F_{2,5}}{2\sqrt{\zeta+1}} & 0 & 0 & 0 \\ 0 & \frac{\zeta(F_{2,3})^* e^{-i(\delta-\phi)}}{2\sqrt{\zeta(\zeta+1)}} & \frac{\zeta}{2\zeta+2} & 0 & \frac{\sqrt{\zeta}e^{i\phi} F_{3,5}}{2\zeta+2} & 0 & 0 & 0 \\ 0 & 0 & 0 & 0 & 0 & 0 & 0 & 0 \\ 0 & \frac{e^{-i\delta}(F_{2,5})^*}{2\sqrt{\zeta+1}} & \frac{\sqrt{\zeta}e^{-i\phi}(F_{3,5})^*}{2\zeta+2} & 0 & \frac{1}{2\zeta+2} & 0 & 0 & 0 \\ 0 & 0 & 0 & 0 & 0 & 0 & 0 & 0 \\ 0 & 0 & 0 & 0 & 0 & 0 & 0 & 0 \\ 0 & 0 & 0 & 0 & 0 & 0 & 0 & 0 \end{pmatrix} \qquad (31)$$

In this case, we initially concentrate our computations solely on subsystem $A$. Writing out the partial transpose for this subsystem yields:
$$[\rho_s(t)]^{T_A}_{W_\zeta} =$$

$$\begin{pmatrix}
0 & 0 & 0 & 0 & 0 & \frac{e^{-i\delta}F_{2,5}{}^*}{2\sqrt{\zeta+1}} & \frac{\sqrt{\zeta}e^{-i\phi}F_{3,5}{}^*}{2\zeta+2} & 0 \\
0 & \frac{1}{2} & \frac{\zeta F_{2,3}e^{i(\delta-\phi)}}{2\sqrt{\zeta(\zeta+1)}} & 0 & 0 & 0 & 0 & 0 \\
0 & \frac{\zeta e^{-i(\delta-\phi)}F_{2,3}{}^*}{2\sqrt{\zeta(\zeta+1)}} & \frac{\zeta}{2\zeta+2} & 0 & 0 & 0 & 0 & 0 \\
0 & 0 & 0 & 0 & 0 & 0 & 0 & 0 \\
0 & 0 & 0 & 0 & \frac{1}{2\zeta+2} & 0 & 0 & 0 \\
\frac{e^{i\delta}F_{2,5}}{2\sqrt{\zeta+1}} & 0 & 0 & 0 & 0 & 0 & 0 & 0 \\
\frac{\sqrt{\zeta}e^{i\phi}F_{3,5}}{2\zeta+2} & 0 & 0 & 0 & 0 & 0 & 0 & 0 \\
0 & 0 & 0 & 0 & 0 & 0 & 0 & 0
\end{pmatrix} \quad (32)$$

because this matrix is Hermitian, its trace norm is equal to the sum of the absolute values of the eigenvalues of the matrix. In accordance with this, the negativity relation is written as follows:

$$N_{A-BC} = -\frac{1}{8(\zeta+1)^{5/2}\sqrt{\zeta(\zeta+1)}}$$

$$(-\zeta\left|\sqrt{\zeta}(\zeta+1)(2\zeta+1) - e^{-i(\delta+\phi)}\sqrt{e^{2i(\delta+\phi)}\zeta(\zeta+1)^2\left(4\zeta(\zeta+1)|F_{2,3}|^2+1\right)}\right|$$

$$-\left|\sqrt{\zeta}(\zeta+1)(2\zeta+1) - e^{-i(\delta+\phi)}\sqrt{e^{2i(\delta+\phi)}\zeta(\zeta+1)^2\left(4\zeta(\zeta+1)|F_{2,3}|^2+1\right)}\right|$$

$$-\zeta\left|\sqrt{\zeta}(\zeta+1)(2\zeta+1) + e^{-i(\delta+\phi)}\sqrt{e^{2i(\delta+\phi)}\zeta(\zeta+1)^2\left(4\zeta(\zeta+1)|F_{2,3}|^2+1\right)}\right|$$

$$-\left|\sqrt{\zeta}(\zeta+1)(2\zeta+1) + e^{-i(\delta+\phi)}\sqrt{e^{2i(\delta+\phi)}\zeta(\zeta+1)^2\left(4\zeta(\zeta+1)|F_{2,3}|^2+1\right)}\right|$$

$$-4\sqrt{\zeta(\zeta+1)}(\zeta+1)^{3/2}\left|\sqrt{(\zeta+1)|F_{2,5}|^2+\zeta|F_{3,5}|^2}\right| + 4\zeta\sqrt{\zeta(\zeta+1)}(\zeta+1)^{3/2}$$

$$+2\sqrt{\zeta(\zeta+1)}(\zeta+1)^{3/2}) \quad (33)$$

With the same calculations as before, we derive the following relationship for the partial transpose of subsystem $B$ in this state:

$$[\rho_s(t)]^{T_B}_{W_\zeta} =$$

$$\begin{pmatrix}
0 & 0 & 0 & \frac{\zeta e^{-i(\delta-\phi)} F_{2,3}{}^*}{2\sqrt{\zeta(\zeta+1)}} & 0 & 0 & \frac{\sqrt{\zeta} e^{i\phi} F_{3,5}}{2\zeta+2} & 0 \\
0 & \frac{1}{2} & 0 & 0 & \frac{e^{i\delta} F_{2,5}}{2\sqrt{\zeta+1}} & 0 & 0 & 0 \\
0 & 0 & \frac{\zeta}{2\zeta+2} & 0 & 0 & 0 & 0 & 0 \\
\frac{\zeta F_{2,3} e^{i(\delta-\phi)}}{2\sqrt{\zeta(\zeta+1)}} & 0 & 0 & 0 & 0 & 0 & 0 & 0 \\
0 & \frac{e^{-i\delta} F_{2,5}{}^*}{2\sqrt{\zeta+1}} & 0 & 0 & \frac{1}{2\zeta+2} & 0 & 0 & 0 \\
0 & 0 & 0 & 0 & 0 & 0 & 0 & 0 \\
\frac{\sqrt{\zeta} e^{-i\phi} F_{3,5}{}^*}{2\zeta+2} & 0 & 0 & 0 & 0 & 0 & 0 & 0 \\
0 & 0 & 0 & 0 & 0 & 0 & 0 & 0
\end{pmatrix} \quad (34)$$

this matrix is also Hermitian and its trace norm equals the sum of the absolute values of its eigenvalues. Furthermore, following the negativity relation, we can express this as:

$$N_{B-CA} = -\frac{1}{8(\zeta+1)^{5/2}\sqrt{\zeta(\zeta+1)}}$$

$$\left(-\zeta \left| \sqrt{\zeta(\zeta+1)(\zeta+2)} - e^{-i(\delta+\phi)}\sqrt{e^{2i(\delta+\phi)}\zeta(\zeta+1)^2(\zeta^2+4(\zeta+1)|F_{2,5}|^2)} \right| \right.$$

$$- \left| \sqrt{\zeta(\zeta+1)(\zeta+2)} - e^{-i(\delta+\phi)}\sqrt{e^{2i(\delta+\phi)}\zeta(\zeta+1)^2(\zeta^2+4(\zeta+1)|F_{2,5}|^2)} \right|$$

$$-\zeta \left| \sqrt{\zeta(\zeta+1)(\zeta+2)} + e^{-i(\delta+\phi)}\sqrt{e^{2i(\delta+\phi)}\zeta(\zeta+1)^2(\zeta^2+4(\zeta+1)|F_{2,5}|^2)} \right|$$

$$- \left| \sqrt{\zeta(\zeta+1)(\zeta+2)} + e^{-i(\delta+\phi)}\sqrt{e^{2i(\delta+\phi)}\zeta(\zeta+1)^2(\zeta^2+4(\zeta+1)|F_{2,5}|^2)} \right|$$

$$-4\sqrt{\zeta}\sqrt{\zeta(\zeta+1)}(\zeta+1)^{3/2}\left|\sqrt{(\zeta+1)|F_{2,3}|^2+|F_{3,5}|^2}\right| + 2\zeta\sqrt{\zeta(\zeta+1)}(\zeta+1)^{3/2}$$

$$\left. +4\sqrt{\zeta(\zeta+1)}(\zeta+1)^{3/2}\right) \quad (35)$$

Finally, for the partial transpose of subsystem $C$, we obtain the following relationship:

$$[\rho_s(t)]_{W_\zeta}^{T_C} =$$

$$\begin{pmatrix}
0 & 0 & 0 & \frac{\zeta F_{2,3} e^{i(\delta-\phi)}}{2\sqrt{\zeta(\zeta+1)}} & 0 & \frac{e^{i\delta} F_{2,5}}{2\sqrt{\zeta+1}} & 0 & 0 \\
0 & \frac{1}{2} & 0 & 0 & 0 & 0 & 0 & 0 \\
0 & 0 & \frac{\zeta}{2\zeta+2} & 0 & \frac{\sqrt{\zeta} e^{i\phi} F_{3,5}}{2\zeta+2} & 0 & 0 & 0 \\
\frac{\zeta e^{-i(\delta-\phi)} F_{2,3}{}^*}{2\sqrt{\zeta(\zeta+1)}} & 0 & 0 & 0 & 0 & 0 & 0 & 0 \\
0 & 0 & \frac{\sqrt{\zeta} e^{-i\phi} F_{3,5}{}^*}{2\zeta+2} & 0 & \frac{1}{2\zeta+2} & 0 & 0 & 0 \\
\frac{e^{-i\delta} F_{2,5}{}^*}{2\sqrt{\zeta+1}} & 0 & 0 & 0 & 0 & 0 & 0 & 0 \\
0 & 0 & 0 & 0 & 0 & 0 & 0 & 0 \\
0 & 0 & 0 & 0 & 0 & 0 & 0 & 0
\end{pmatrix} \quad (36)$$

it can be shown that this matrix is also Hermitian, and its trace norm equals the sum of the absolute values of its eigenvalues. Furthermore, following the negativity relation, we can express:

$$N_{C-AB} = -\frac{1}{8(\zeta+1)^{3/2}\sqrt{\zeta(\zeta+1)}}$$
$$(-\left|\sqrt{\zeta}(\zeta+1)^2 - e^{-i(\delta+\phi)}\sqrt{e^{2i(\delta+\phi)}\zeta(\zeta+1)^2\left(4\zeta|F_{3,5}|^2 + (\zeta-2)\zeta + 1\right)}\right|$$
$$-\left|\sqrt{\zeta}(\zeta+1)^2 + e^{-i(\delta+\phi)}\sqrt{e^{2i(\delta+\phi)}\zeta(\zeta+1)^2\left(4\zeta|F_{3,5}|^2 + (\zeta-2)\zeta + 1\right)}\right|$$
$$-4\zeta\sqrt{\zeta(\zeta+1)}\left|\sqrt{\zeta|F_{2,3}|^2 + |F_{2,5}|^2}\right| - 4\sqrt{\zeta(\zeta+1)}\left|\sqrt{\zeta|F_{2,3}|^2 + |F_{2,5}|^2}\right|$$
$$+2\sqrt{\zeta(\zeta+1)}(\zeta+1)^{3/2}) \tag{37}$$

## IV Results and discussion

In this particular part of our study, we would like to show case and highlight the detailed computational findings and outcomes pertaining to the $W_\zeta$ state being employed as the initial condition for the triple-qubit system analysis. As previously mentioned, we have utilized the negativity criterion to analyze the dynamics of quantum entanglement.

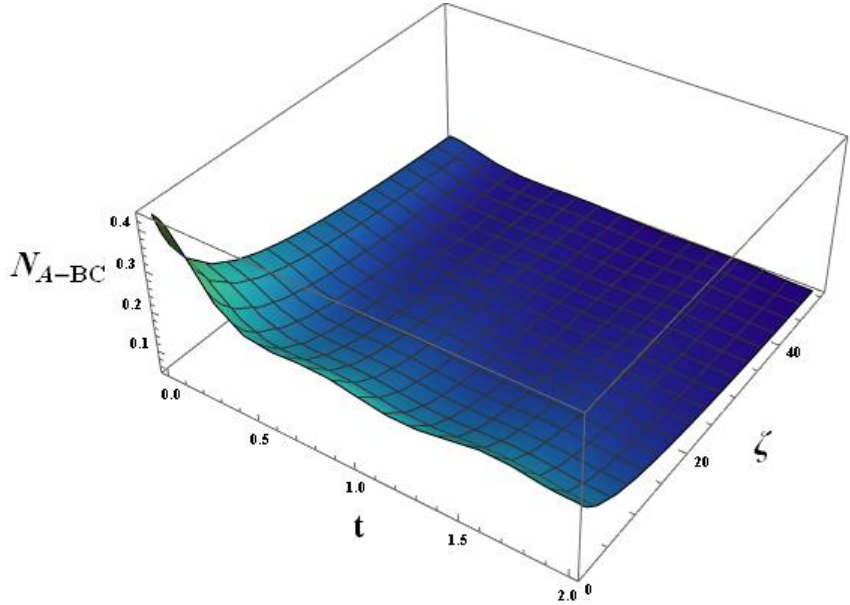

Figure 1: The 3D diagram of $N_{A-BC}$ for state $W_\zeta$, with respect to $\zeta$ and $t$ for specific values of $\delta = \pi/2$, $\phi = \pi/2$, $\eta = 1$, $T = 0.5$, $\gamma = 1$, $\alpha = 1$, $g_A = 0.1$, $g_B = 0.2$, $g_C = 0.3$, and $\mathcal{N} = 51$.

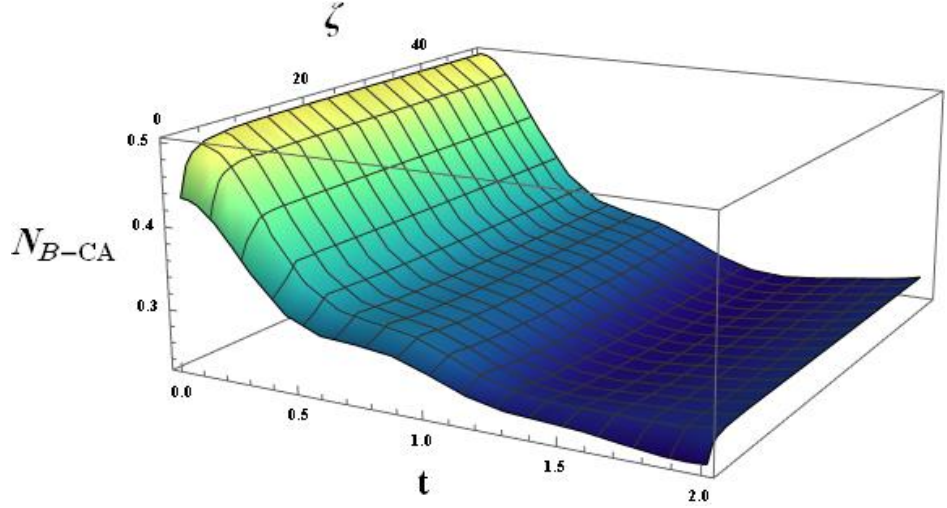

Figure 2: The 3D plot of $N_{B-CA}$ for state $W_\zeta$, with respect to $\zeta$ and $t$ for specific values of $\delta = \pi/2$, $\phi = \pi/2$, $\eta = 1$, $T = 0.5$, $\gamma = 1$, $\alpha = 1$, $g_A = 0.1$, $g_B = 0.2$, $g_C = 0.3$, and $\mathcal{N} = 51$.

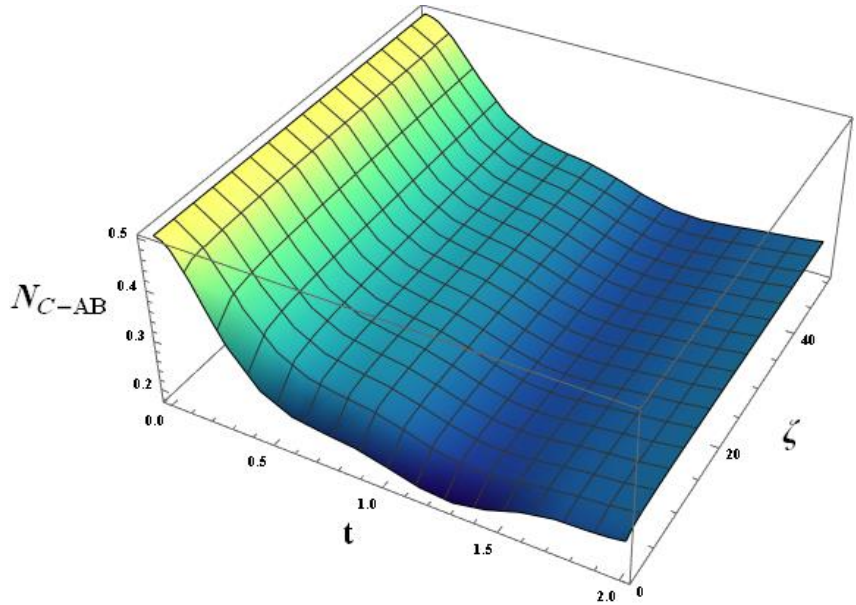

Figure 3: The 3D diagram of $N_{C-AB}$ for the $W_\zeta$ state, with respect to $\zeta$ and $t$ for specific values of $\delta = \pi/2$, $\phi = \pi/2$, $\eta = 1$, $T = 0.5$, $\gamma = 1$, $\alpha = 1$, $g_A = 0.1$, $g_B = 0.2$, $g_C = 0.3$, and $\mathcal{N} = 51$.

Figure 1 shows the three-dimensional diagram of $N_{A-BC}$ for state $W_\zeta$, with respect to $\zeta$ and $t$. This figure illustrates the trend of $N_{A-BC}$ approaching zero with some fluctuation as time progresses. As $\zeta$ increases, $N_{A-BC}$ decreases. Initially, the decrease in $N_{A-BC}$ is less pronounced due to increasing $\zeta$. This instance shows the quickest decline in negativity as time and $\zeta$ increase and we see the death of entanglement from $t = 0.3$ onwards.

Figure 2 depicts the three-dimensional diagram of $N_{B-CA}$ for state $W_\zeta$, with respect to $\zeta$ and $t$. This figure demonstrates the decrease in $N_{B-CA}$ over time. As $\zeta$ varies from 0 to 5, $N_{B-CA}$ increases from 0 to 0.15. The negativity changes in this scenario exhibit sluggish progression with rising $\zeta$ values, however, with the passage of time, starting from $t = 1.3$, entanglement swiftly diminishes.

Figure 3 shows the three-dimensional diagram of $N_{C-AB}$ for the $W_\zeta$ state, with respect to $\zeta$ and $t$. This figure shows that the fluctuation of $N_{C-AB}$ ranges from 0 to 1.25 over time, reaching 0 at $t = 1.25$. Subsequently, $N_{C-AB}$ rises to 0.22 between $t = 1.25$ and $t = 2$. With increasing $\zeta$, $N_{C-AB}$ escalates gradually from 0 to 5, hovering between 0 and 0.25, and maintaining near-constant values between 5 and 50. As depicted in Figure 2, the alterations in negativity for this scenario also exhibit a gradual pace with increasing $\zeta$ values. However, as time progresses, starting from $t = 1.3$ onwards, entanglement experiences a rapid decline.

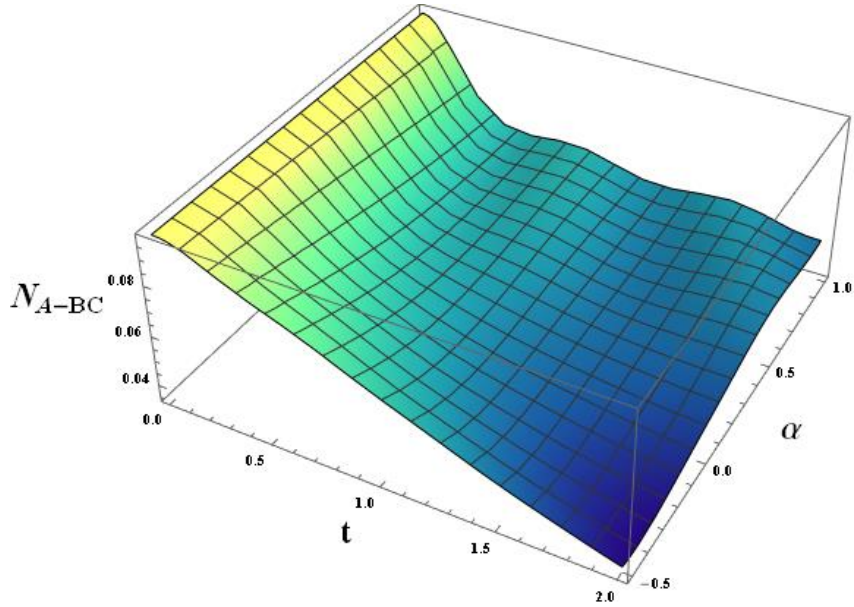

Figure 4: The 3D diagram of $N_{A-BC}$ for $W_\zeta$ state, in relation to $\alpha$ and $t$ for specific values of $\delta = \pi/2$, $\phi = \pi/2$, $\eta = 1$, $T = 0.5$, $\gamma = 1$, $\zeta = 50$, $g_A = 0.1$, $g_B = 0.2$, $g_C = 0.3$, and $\mathcal{N} = 51$.

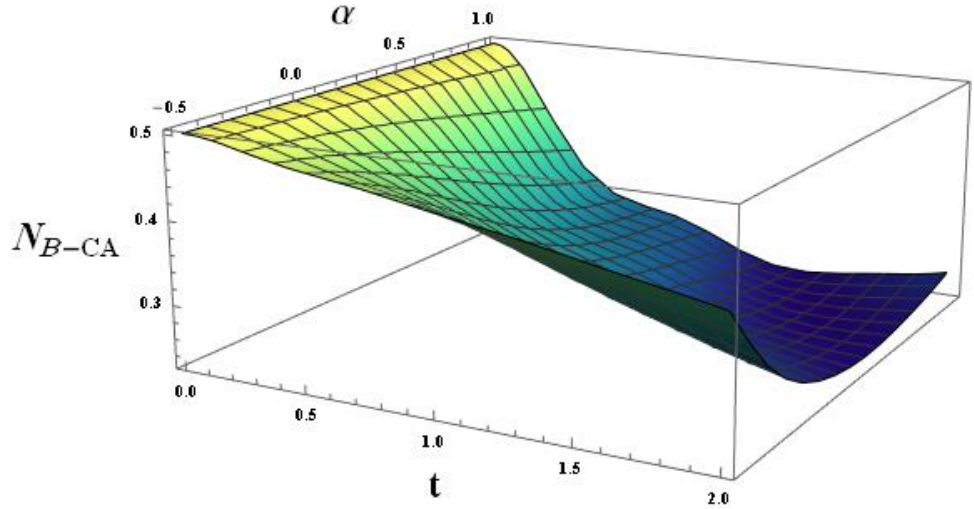

Figure 5: The 3D diagram of $N_{B-CA}$ for state $W_\zeta$, in relation to $\alpha$ and $t$ for specific values of $\delta = \pi/2$, $\phi = \pi/2$, $\eta = 1$, $T = 0.5$, $\gamma = 1$, $\zeta = 50$, $g_A = 0.1$, $g_B = 0.2$, $g_C = 0.3$, and $\mathcal{N} = 51$.

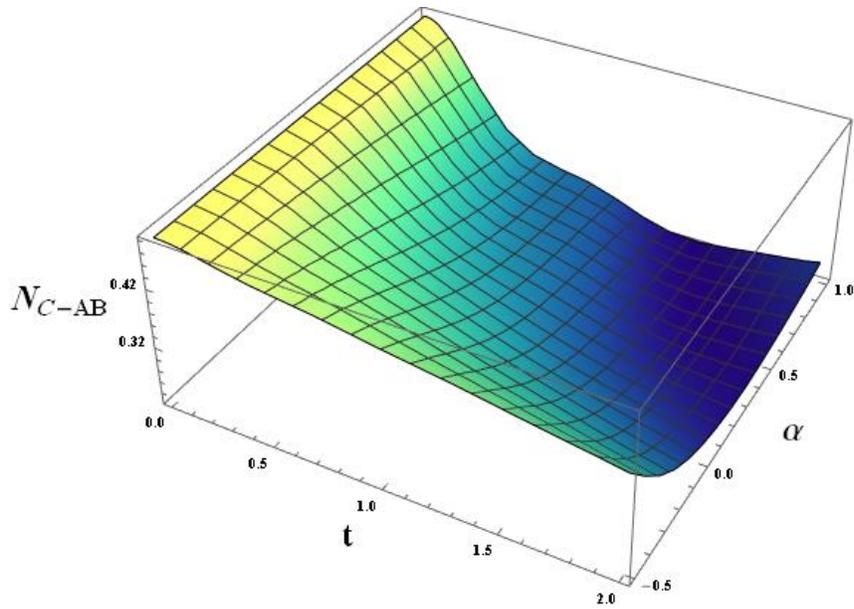

Figure 6: The 3D diagram of $N_{C-AB}$ for state $W_\zeta$, in relation to $\alpha$ and $t$ for specific values of $\delta = \pi/2$, $\phi = \pi/2$, $\eta = 1$, $T = 0.5$, $\gamma = 1$, $\zeta = 50$, $g_A = 0.1$, $g_B = 0.2$, $g_C = 0.3$, and $\mathcal{N} = 51$.

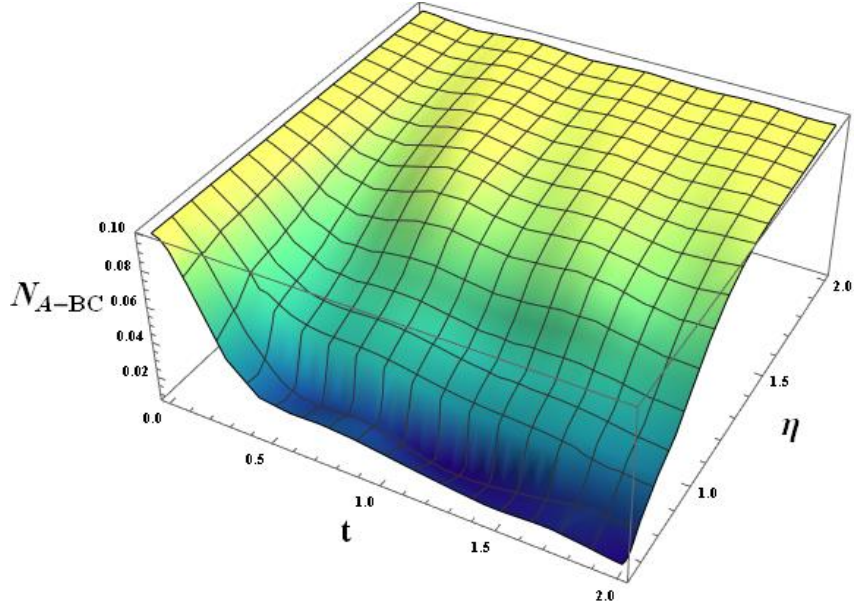

Figure 7: The 3D diagram of $N_{A-BC}$ for state $W_\zeta$, in relation to $\eta$ and $t$ for specific values of $\delta = \pi/2$, $\phi = \pi/2$, $\alpha = 1$, $T = 0.5$, $\gamma = 1$, $\zeta = 50$, $g_A = 0.1$, $g_B = 0.2$, $g_C = 0.3$, and $\mathcal{N} = 51$.

Figure 4 shows the three-dimensional diagram of $N_{A-BC}$ for $W_\zeta$ state, in relation to $\alpha$ and $t$. This figure depicts the decline of $N_{A-BC}$ with time. Decreases in $\alpha$ accelerate $N_{A-BC}$ reduction over time. Conversely, as $\alpha$ rises, the rate of $N_{A-BC}$ decline reduces. Transitioning from -0.5 to 1, $N_{A-BC}$ ascends from 0 to 0.05. For $\alpha > 0$, $N_{A-BC}$ exhibits minor fluctuations over time. In this scenario, with $\alpha = 0$, entanglement diminishes linearly as time progresses. As $\alpha$ increases, the rate of entanglement reduction slows down over time.

Figure 5 depicts the three-dimensional diagram of $N_{B-CA}$ for state $W_\zeta$, in relation to $\alpha$ and $t$. This figure shows the reduction of $N_{B-CA}$ unfolds over time. Sluggish changes occur with lower $\alpha$ values, while increments in $\alpha$ correlate with decreases in $N_{B-CA}$ until it reaches a minimum of 0.22 at $\alpha = 0.3$. Subsequently, an upward trend in $N_{B-CA}$ commences, peaking at 0.26 for $\alpha = 1$. Here as well, at $\alpha = 0$, entanglement decreases linearly over time, albeit with a slightly gentler slope compared to the scenario depicted in Figure 4. As $\alpha$ increases, the process of entanglement reduction speeds up over time.

Figure 6 shows the three-dimensional diagram of $N_{C-AB}$ for state $W_\zeta$, in relation to $\alpha$ and $t$. This figure shows the declining trend of $N_{C-AB}$ over time. Similar to $N_{B-CA}$, $N_{C-AB}$ varies inversely with $\alpha$, reaching a minimum of 0.24 at $\alpha = 0.35$, then ascending to 0.28 at $\alpha = 1$. In this instance, the entanglement at $\alpha = 0$ diminishes gradually over time. As $\alpha$ increases, the speed of entanglement reduction also increases.

Figure 7 depicts the three-dimensional diagram of $N_{A-BC}$ for state $W_\zeta$, in relation to $\eta$ and $t$. This figure shows that $N_{A-BC}$ decreases within the time interval between $\eta = 0.65$ and $t = 0.5$ with slight fluctuations. The increase from $\eta = 0.65$ to 2 results in $N_{A-BC}$ escalating from 0 to 1, more rapidly at lower times. Between $\eta = 1$ and 2, temporal fluctuations in $N_{A-BC}$ emerge. In this scenario, the entanglement at $\eta = 0.65$ decreases quickly over time. As $\eta$ increases, the entanglement also increases and reaches its maximum value of 0.1 at $\eta = 2$.

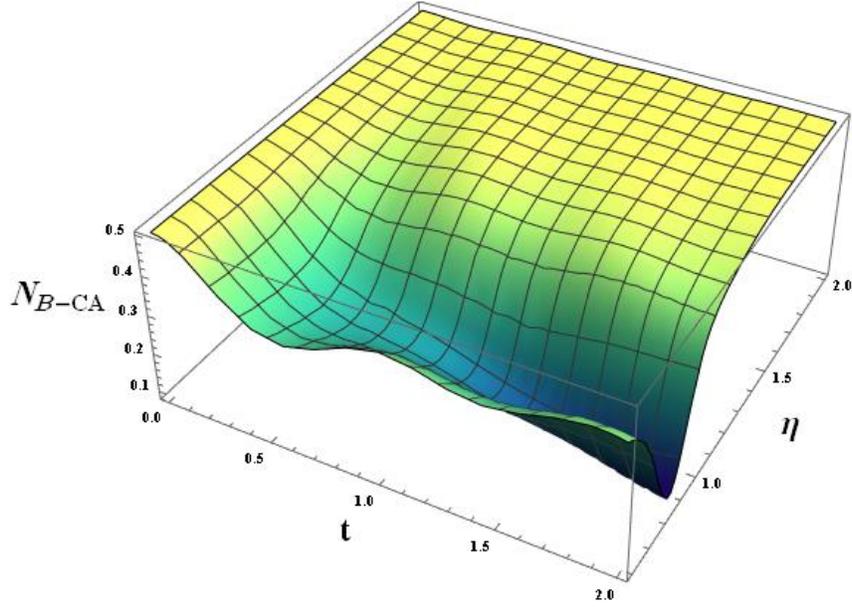

Figure 8: The 3D diagram of $N_{B-CA}$ for state $W_\zeta$, in relation to $\eta$ and $t$ for specific values of $\delta = \pi/2$, $\phi = \pi/2$, $\alpha = 1$, $T = 0.5$, $\gamma = 1$, $\zeta = 50$, $g_A = 0.1$, $g_B = 0.2$, $g_C = 0.3$, and $\mathcal{N} = 51$.

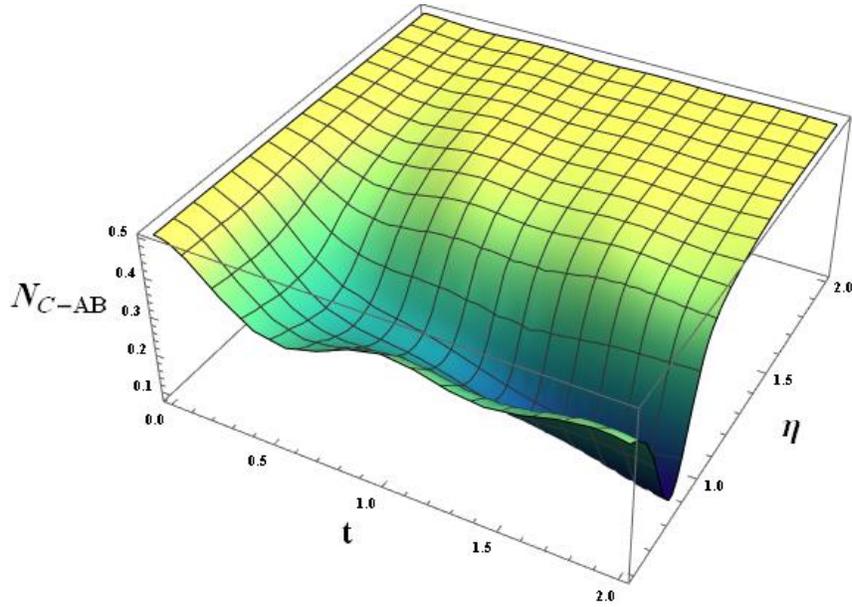

Figure 9: The 3D diagram of $N_{C-AB}$ for the $W_\zeta$ state, with respect to $\eta$ and $t$ for specific values of $\delta = \pi/2$, $\phi = \pi/2$, $\alpha = 1$, $T = 0.5$, $\gamma = 1$, $\zeta = 50$, $g_A = 0.1$, $g_B = 0.2$, $g_C = 0.3$, and $\mathcal{N} = 51$.

Figure 8 shows the three-dimensional diagram of $N_{B-CA}$ for state $W_\zeta$, in relation to $\eta$ and $t$. This figure shows cases alternating fluctuations in $N_{B-CA}$ at $\eta = 0.65$, with local minima at $t = 0.6$ (0.32) and $t = 1.4$ (0.37). As $\eta$ climbs from 0.65 to 0.74, $N_{B-CA}$ declines, hitting 0.1 at $\eta = 0.74$. Subsequently, a rapid ascent occurs between $\eta = 0.74$ and $\eta = 1.2$, with a more gradual increase to 0.5 from 1.2 to 2. In this scenario, for the $\eta = 0.65$, the entanglement decreases slightly over time. With the increase of $\eta$ from 0.65 to 0.74, the process of entanglement reduction accelerates quickly over time. From $\eta = 0.74$ to $\eta = 1.2$, the entanglement increases swiftly. From $\eta = 1.2$ to $\eta = 2$, the entanglement increases gradually.

Figure 9 depicts the three-dimensional diagram of $N_{C-AB}$ for the $W_\zeta$ state, with respect to $\eta$ and $t$. This figure shows that $N_{C-AB}$ displays similar fluctuations at $\eta = 0.65$, reaching local minima at $t = 0.6$ (0.31) and $t = 1.3$ (0.33). The negativity reduces from $\eta = 0.65$ to 0.78, hitting 0.1 at $\eta = 0.78$. Between $\eta = 0.78$ and 1.2, $N_{C-AB}$ rises rapidly before gradually reaching 0.5 between 1.2 and 2. In this case, as in Figure 8, the discussions raised about Figure 8, for the dynamics of quantum entanglement, are still valid. Based on the information presented, the quantum entanglement of the three-qubit system $W_\zeta$ appears to mainly decrease for the parameters $\zeta$ and $\alpha$, increase for the parameter $\eta$.

Quantum entanglement is a fascinating phenomenon in quantum mechanics in which particles become interconnected in such a way that the quantum state of one particle depends on the state of another, no matter how far apart they are from one another in space. This remarkable property has significant implications, leading to various applications across various fields, including technology and scientific research. Quantum negativity serves as a crucial measure of entanglement specifically within a mixed bipartite quantum system, providing insights into the degree of entanglement present. A higher degree of negativity typically signifies that the entanglement is stronger. According to the results obtained, it has been observed that the stability of entanglement, often denoted as negativity, for the $B - CA$ and $C - AB$ subsystems is notably superior when compared to the A-BC subsystem. As $\alpha$ increases, the entanglement stability of the $A - BC$ subsystem improves relative to the $B - CA$ and $C - AB$ subsystems. As the value of $\eta$ increases over time, we observe that the stability of entanglement progressively rises for each of the three subsystems under consideration. This indicates a positive correlation between the increase in $\eta$ and the enhancement of entanglement stability across all three subsystems. Therefore, a positive relationship between increasing η and increasing entanglement stability is seen.

In the past, investigations on the dynamics of quantum entanglement concerning three-qubit states were conducted by Kazemi et al. [18] and Jafarpour et al. [19], in addition to several other scholarly works on the subject matter. Even though the quantum states analyzed in these publications vary from the $W_\zeta$ quantum state scrutinized in our research, the results generated from our calculations demonstrate a strong coherence with the findings outlined in the previously mentioned articles. Nevertheless, the theoretical and experimental investigation into the $W_\zeta$ state has yet to be explored until the publication of this particular article. The entanglement of a quantum state is inherently fragile and sensitive, as it is subjected to inevitable couplings with various environmental degrees of freedom. These couplings lead to the dissipative evolution of quantum coherence, resulting in the gradual loss of useful entanglement that occurs during both particle propagation and computational processes. This phenomenon, often referred to as decoherence, represents a significant and formidable challenge in the quest to build practical and reliable quantum computers. No matter how much shielding and protective measures are implemented, quantum devices still find themselves vulnerable to decoherence and decay processes, which can severely hinder their operational effectiveness. Consequently, developing a comprehensive understanding of disentanglement within the specific system of interest becomes crucial, not only from a theoretical standpoint but also for experimental validation. This understanding is essential to devise strategies that can mitigate the effects of decoherence and maintain the integrity of quantum entanglement over time, thus enhancing the viability and performance of future quantum computing technologies. By investigating the mechanisms and factors that contribute to disentanglement, researchers can better address the challenges posed by decoherence, making strides toward more robust quantum systems that can function effectively in real-world applications.

## V  Conclusions

The study aims to understand how entanglement evolves during interactions between quantum systems. This knowledge is crucial for improving our understanding of quantum behaviour, which can contribute to advancements in quantum computing and information processing. By examining the entanglement properties in this context, researchers seek to uncover fundamental principles of quantum mechanics and explore practical applications of entangled states in technology.

In this particular research endeavor, we delved into the examination of the intricate quantum entanglement dynamics exhibited by the $W_\zeta$ state using the innovative Hamiltonian put forth in this scholarly work. The preliminary states taken into account are those characterized as $W_\zeta$. Our insightful discoveries shed light on the phenomenon where entanglement experiences a decline with the amplification of the $XZY - YZX$ triplet interaction's potency, and concurrently with the mounting of the $\zeta$ parameter. Interestingly, however, there is a contrasting increase in entanglement when considering elevated values of the strength emanating from the externally applied transverse field directed towards the surrounding environment. The extensive computations undertaken in this study harmoniously align with the wealth of prior literature tackling the entanglement dynamics pertaining to a select state configuration within the enveloping $W_\zeta$ state, specifically denoted as $W$ (a nuanced illustration of the $W_\zeta$ state with $\zeta = 1$, $\delta = 0$, and $\phi = 0$). The results show that the stability of entanglement is significantly higher in the $B - CA$ and $C - AB$ subsystems compared to the $A - BC$ subsystem. As $\alpha$ increases, the stability of the $A - BC$ subsystem improves relative to the others. Additionally, as $\eta$ increases over time, entanglement stability rises in all three subsystems, indicating a positive correlation between $\eta$ and entanglement stability. Thus, there is a clear relationship between increasing $\eta$ and enhancing entanglement stability.